%% file: Hoefener00EPL.tex
\documentstyle[epsf]{europhys}
\input euromacr

\begin{document}

\euro{--} {--} {--} {accepted for publication, 04/2000}
\Date{April 2000} \shorttitle{C. H\"ofener et al.,  Voltage and
temperature dependence of the magnetoresistance in manganites}

\title{Voltage and temperature dependence of the grain \linebreak boundary tunneling magnetoresistance
in manganites}

\author{\renewcommand{\thefootnote}{\alph{footnote})}
C.~H\"{o}fener, J.B.~Philipp, J.~Klein, L.~Alff, A.~Marx, B.
B\"{u}chner, and R.~Gross}

\institute{II.~Physikalisches Institut, \\ Universit\"{a}t zu
K\"{o}ln, Z\"{u}lpicherstr.~77, D - 50937 K\"{o}ln, Germany
     }
\rec{30 November 1999}{5 April 2000}

 \pacs{
 \Pacs{75}{70.Cn}{Interfacial magnetic properties}
 \Pacs{75}{70.Ak}{Magnetic properties of monolayers and thin films}
 \Pacs{73}{40.-c}{Electronic transport in interface structures}
     }

\maketitle

\begin{abstract}

We have performed a systematic analysis of the voltage and temperature
dependence of the tunneling magnetoresistance (TMR) of grain boundaries (GB)
in the manganites. We find a strong decrease of the TMR with increasing
voltage and temperature. The decrease of the TMR with increasing voltage
scales with an increase of the inelastic tunneling current due to multi-step
inelastic tunneling via localized defect states in the tunneling barrier.
This behavior can be described within a three-current model for magnetic
tunnel junctions  that extends the two-current Julli\`{e}re model by adding an
inelastic, spin-independent tunneling contribution.  Our analysis gives
strong evidence that the observed drastic decrease of the GB-TMR in
manganites is caused by an imperfect tunneling barrier.

\end{abstract}

The tunneling resistance between two ferromagnetic metal layers
separated by a thin insulating barrier depends on the relative
orientation of the magnetization and the electron spin
polarization in each layer \cite{Julliere:75a}. Since for
materials with large spin polarization a large tunneling
magnetoresistance (TMR) can be achieved, magnetic tunnel junctions
have recently attracted much attention and their use in memory
devices in envisaged \cite{Moodera:95a,Parkin:99a}. Due to their
half-metallic ferromagnetic state with only a single spin band
crossing the Fermi level, the spin polarization of the perovskite
manganites of composition $\rm La_{2/3}{\it D}_{1/3}MnO_3$ with
$D=$ Ca, Sr, and Ba is close to 100\%. According to the Julli\`{e}re
model\cite{Julliere:75a}, a high TMR ratio $\Delta R/R=
(R_{\uparrow\downarrow}
-R_{\uparrow\uparrow})/R_{\uparrow\uparrow} =(2P_1P_2)/(1-P_1P_2)$
is expected making these materials attractive for magnetic tunnel
junctions. Here, $R_{\uparrow\uparrow}$ and
$R_{\uparrow\downarrow}$ is the tunneling resistance for parallel
and anti-parallel magnetization orientation and $P_i=2a_i-1$ is
the spin polarization where $a_i$ is the fraction of majority spin
electrons in the density of states of layer $i$. Indeed a high TMR
ratio well above 100\% has been achieved at low temperatures and
low applied fields $H$ of the order of 10\,mT
\cite{Viret:97a,Obata:99a}. In addition to planar type tunnel
junctions, a large low field magnetoresistance was found for grain
boundaries (GBs) in the perovskite manganites
\cite{Gubkin:93a,Hwang:96a,Gupta:96a,Mathur:97a,Klein:99a,Gross:99a}.
In contrast to the colossal magnetoresistance (CMR) observed in
single crystals, the GB magnetoresistance can be observed at low
$H$ ($\sim 10$\,mT) and over the entire temperature range below
the Curie temperature $T_C$. The GB magnetoresistance is
attributed to spin-polarized tunneling across an insulating GB
barrier between two ferromagnetic grains
\cite{Klein:99a,Gross:99a,Ziese:99a,Lee:99a}. For both planar
junctions and grain boundary junctions (GBJs) based on the
manganites a large TMR has been observed only at low temperatures
and junction voltages, which strongly decreases with increasing
temperature and bias voltage \cite{Viret:97a,Obata:99a}. The
origin of this effect has not been clarified so far. A similar but
somewhat weaker decrease of the TMR also has been found for
magnetic tunnel junctions employing Al$_2$O$_3$ barriers and Co
and Permalloy electrodes\cite{Marley:97a,Sousa:99a} as well as for
granular\cite{Berkowitz:92a} and powder systems \cite{Coey:99a}.
Also for the latter systems the origin of the voltage and
temperature dependence of the TMR is discussed controversially.

In this Letter we report on the systematic study of the GB
magnetoresistance in \linebreak[4] $\rm La_{2/3}Ca_{1/3}MnO_3$
(LCMO) as a function of voltage $V$ and temperature $T$ using well
defined single GBJs. By analyzing the current-voltage
characteristics (IVCs) in terms of the Glazman-Matveev (GM)
theory\cite{Glazman:88a} for multi-step tunneling we show that
with increasing $V$ inelastic tunneling contributes more and more
to the charge transport across the GB barrier. Since spin
polarization may be not conserved in the inelastic channel, the
inelastic tunneling current does not contribute to the
magnetoresistance and the measured TMR ratio is reduced
accordingly. In the presence of the inelastic channel, the well
known two-current model of Julli\`{e}re\cite{Julliere:75a}, which
assumes that the two spin species of electrons tunnel elastically,
has to be extended to a three-current model to account for the
inelastic, spin-independent tunneling current. Within such
three-current model the TMR ratio $\Delta R/R$ is shown to be
proportional to the ratio $I^{e}/(I^{e}+I^{i})$  of the elastic
tunneling current, $I^{e}$, and the sum of the elastic and
inelastic tunneling current, $(I^{e}+I^{i})$ as discussed in
detail below. Such proportionality has been found in our
experiments on LCMO-GBJs.

\begin{figure}[h]
\centerline{\mbox{\epsfxsize=0.65\columnwidth
\epsffile{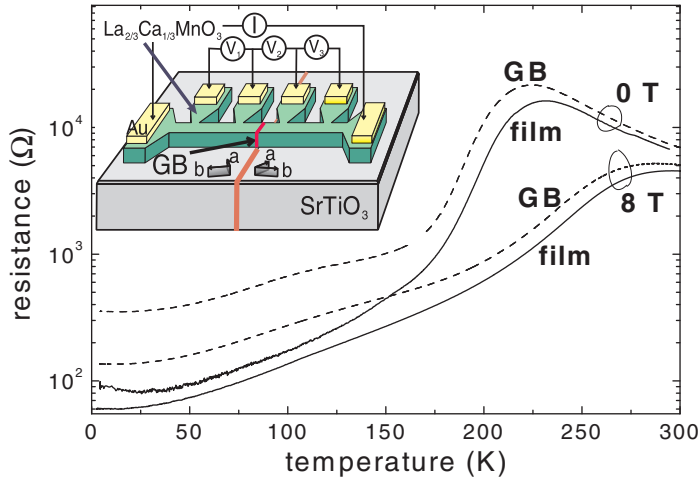}}} \caption[]{\small Resistance
vs. temperature of a 80\,nm thick
La$_{2/3}$Ca$_{1/3}$MnO$_{3-\delta}$ thin film with (dashed lines)
and without (solid lines) grain boundary at zero field and at
8\,T. All curves were measured at a bias voltage of $V=10$\,mV.
The inset shows a sketch of the sample configuration. }
 \label{rt}
\end{figure}

The LCMO-GBJs were fabricated by pulsed laser deposition of epitaxial, 20 to
100\,nm thick LCMO films on symmetrical, [001] tilt SrTiO$_3$ bicrystals
with a misorientation angle of 24$^{\circ}$
\cite{Gross:99a,Wiedenhorst:99a}. After film deposition the samples were
annealed  at 950$^{\circ}$C in oxygen atmosphere for one hour. X-ray
analysis of the films showed only (00$\ell$) reflexes and the FWHM of the
(002) rocking curve before annealing typically was $\le0.03^{\circ}$. After
the annealing process the FWHM slightly increased but stayed below
0.1$^{\circ}$. Microbridges of 10 to 30\,$\mu$m width straddling the GB were
patterned using optical lithography and Ar ion beam etching. In this way
well defined individual GBJs are obtained. For comparison we also fabricated
microbridges of exactly the same size that do not cross the GB as shown in
the inset of Fig.~\ref{rt}. These microbridges without GB are used for the
analysis of the film properties. They also are used to determine the
additional voltage drop along the film adjacent to the GB. This voltage drop
can then be subtracted from the total voltage measured for an identical
microbridge with GB to get the voltage drop across the GB alone. In the same
way, the small series resistance due to the film can be determined and
subtracted from the measured total resistance to obtain the pure GB
resistance and TMR ratio. Further details on the transport properties and
the microstructure of the GBJs have been reported recently
\cite{Klein:99a,Gross:99a}.


\vspace*{0cm}


Typical resistance vs temperature, $R(T)$, curves of a LCMO film are shown
in Fig.~\ref{rt}. The maximum in the $R(T)$ curve that can be associated
with $T_C$ is about 225\,K  and is shifted to 275\,K at 8\,T. Comparing the
$R(T)$ curves of microbridges of the same size but with and without GB, the
resistance of those containing a single GB is found to be enhanced
considerably below $T_C$ as has been discussed in detail
elsewhere\cite{Klein:99a,Gross:99a}. Below about 160\,K the resistance of
bridges with GB is dominated by the GB resistance. The appearance of an
additional GB resistance below $T_C$ that becomes dominant at $T\ll T_C$
recently has been discussed in terms of the formation of a depletion layer
at a disordered, paramagnetic GB interface resulting in a tunneling
barrier\cite{Klein:99a,Gross:99a}.


\vspace*{0cm}
\begin{figure}[h]
\centerline{\mbox{\epsfxsize=0.65\columnwidth
\epsffile{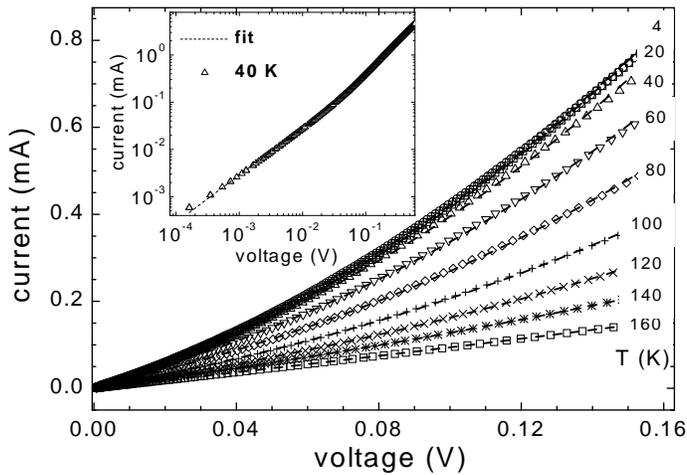}}} \caption[]{\small
Current-voltage characteristics of a 24$^{o}$ [001] tilt GBJ in a
80\,nm thick La$_{2/3}$Ca$_{1/3}$MnO$_{3-\delta}$ film. The dashed
lines are fits to the Glazman-Matveev model. The inset shows the
IVC for $T=40$\,K on a log-log scale.}
 \label{iu}
\end{figure}

 In Fig.~\ref{iu} typical IVCs of a LCMO-GBJ are shown between 4
and 160\,K. In order to get the pure characteristics of the GB we
have corrected the measured data by subtracting the additional
series resistance of the film adjacent to the GB. The film
resistance, which is smaller than about 30\% of the GB resistance
for $4\le T\le 160$\,K, has been determined from an identical
microbridge without GB on the same substrate. Whereas the IVCs of
microbridges without GB are ohmic, highly non-linear IVCs are
found for the microbridges containing a GB with the non-linearity
increasing with decreasing $T$. We also note, that linear IVCs are
found for $T\ge T_C$. It has been shown by Klein {\it et
al.}\cite{Klein:99a,Gross:99a} that the non-linear IVCs of
bicrystal GBJs in the manganites can be well fitted by the
GM-theory\cite{Glazman:88a}. In this theory, in addition to the
elastic tunneling channels multi-step tunneling via a number of
$n$ localized states within the tunneling barrier is taken into
account. Within the GM-model the conductivity $G(V,T)$ is given by
\begin{equation}
G(V,T)=G_0 + \sum_{n=1}^{\infty} G_n (V,T),
\end{equation}
where the conductance $G_0$ represents the direct tunneling term and $G_n$
the tunneling via $n\ge 1$ localized states. For $eV \ll k_BT$ and $eV \gg
k_BT$, $G_n(V,T)$ can be expressed as
\begin{eqnarray}
G_n(V) &=& a_n \cdot V^{(n-\frac{2}{n+1})} \;\;\; {\rm for} \;\; eV \gg k_BT
\\
G_n(T) &=& b_n \cdot T^{(n-\frac{2}{n+1})} \;\;\;\, {\rm for} \;\; eV \ll
k_BT,
\end{eqnarray}
where $a_n$, $b_n \propto {\rm exp}(-2d/(n+1)\alpha )$ are constants
depending on the radius $\alpha$ of the localized states, their density and
the barrier thickness $d$. We note that these expressions are valid as long
as $eV$ and $k_BT$ are small compared to the barrier height. That is, with a
barrier height of the order of 1\,eV\cite{Klein:99a,Gross:99a} the
GM-expressions can be used over the entire $T$ range of our experiments and
for voltages up to about 100\,mV. We note that $eV\gg k_BT$ for most part of
the measured IVCs at $T\le 160$\,K. Then, the IVCs can be fitted by
\begin{equation}
I=G_0 V + a_1 V + a_2 V^\frac{7}{3} + a_3 V^\frac{7}{2} + \cdots \;\; .
\end{equation}
According to the GM-theory, direct tunneling and tunneling via a single
localized state ($n=1$) gives the elastic tunneling current $I^{e}$, whereas
the $n\ge 2$ channels yield the inelastic tunneling current $I^{i}$. Fitting
the IVCs at different $T$ to the GM-theory we can derive the ratio
$I^{e}/(I^{e}+I^{i})$ as a function of $V$ and $T$.   As demonstrated by
Fig.~2, the experimental data can be well fitted by the GM-expressions at
all $T$ up to $V\simeq 0.15$\,V. For all samples the coefficients $a_i$ with
$i\ge 4$ were negligible, i.e. only inelastic channels up to $n=3$ are
required to fit the data. The fits cover more than three orders of magnitude
on the current and voltage axis as shown in the inset of Fig.~\ref{iu}. This
gives strong evidence that the non-linear IVCs are caused by inelastic
tunneling processes which increase with increasing $V$. We note that not all
samples could be fitted as perfect as shown in Fig.~\ref{iu}. However, the
deviations from the GM-model always were small and fits to other models,
e.g. the Simmons model\cite{Simmons:63a} gave much worse results. Deviations
from the GM-theory are expected in the presence of additional inelastic
effects such as the excitation of surface magnons\cite{Guinea:98a} that are
not included in GM theory.

We now discuss the expected interdependence between the ratio
$I^{e}/(I^{e}+I^{i})$ and the measured TMR ratio $\Delta R/R$. To take into
account the inelastic tunneling current we use a three-current model that
extends the well known two-current Julli\`{e}re model\cite{Julliere:75a} by
adding inelastic tunneling channels. In the following we assume that the
inelastic tunneling current depends on $V$ and $T$ but is the same for the
parallel and anti-parallel magnetization arrangement, i.e $I^{i}_{ap} =
I^{i}_{p} = I^{i}$. This is in contrast to the $V$ and $T$ independent
elastic tunneling current which, however, is spin-dependent resulting in
$I^{e}_{ap} < I^{e}_{p}$. The assumption that $I^{i}$ does not depend on the
relative magnetization orientation is justified, since spin orientation is
expected to be not conserved in an inelastic tunneling process. With this
assumption the three-current model yields
\begin{equation}
\frac{\Delta R}{R}(V,T) = \frac{I^{e}_{ap}}{I^{e}_{ap} + I^{i}}(V,T) \left(
\frac{I^{e}_{p}}{I^{e}_{ap}} - 1 \right) .
\end{equation}
It can further be shown that the expression in brackets is given by the $V$
independent TMR ratio of the Julli\`{e}re model resulting in
\begin{equation}
\frac{\Delta R}{R}(V,T) = \frac{I^{e}_{ap}}{I^{e}_{ap} + I^{i}}(V,T) \left(
\frac{\Delta R}{R} \right)_ {\rm Julli\grave{e}re}(T) .
\end{equation}
That is, the TMR ratio $\frac{\Delta R}{R}(V)$ measured at constant
temperature is expected to be proportional to $\frac{I^{e}_{ap}}{I^{e}_{ap}
+ I^{i}}(V)$. In order to check such possible proportionality we have
measured the TMR ratio as a function of $V$ for different $T$. The ratio
$\frac{I^{e}_{ap}}{I^{e}_{ap} + I^{i}}(V)$ has been determined independently
from the zero field IVCs measured at constant $T$ as discussed above.

\vspace*{0cm}
\begin{figure}[h]
\centerline{\mbox{\epsfxsize=0.65\columnwidth
\epsffile{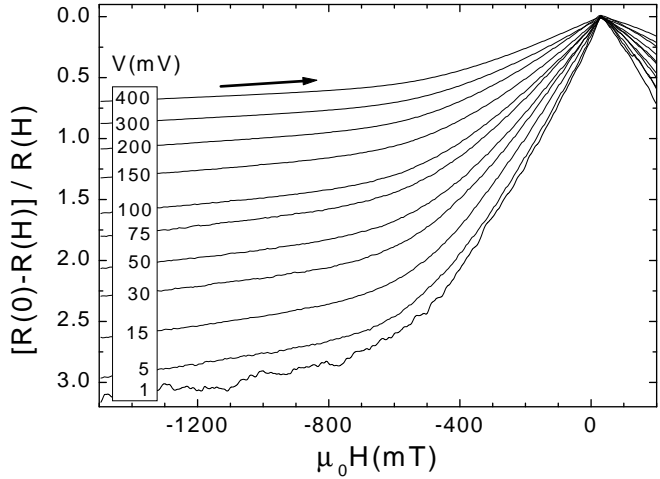 }}} \caption[]{\small
$R(0)-R(H)/R(H)$ measured at different bias voltage plotted versus
the applied magnetic field at 4\,K for a 24$^{o}$ [001] tilt
LCMO-GBJ. The arrow indicates the direction of the field sweep.
The field is applied within the film plane parallel to the
current.  }
 \label{rh}
\end{figure}


In Fig.~\ref{rh}, we have plotted $[R(0)-R(H)]/R(H)$ versus the applied
magnetic field measured at different bias voltages. For the bias voltage
only the voltage drop across the GB is used. In order to determine the TMR
ratio, for $R_{\uparrow\downarrow}$ the maximum of the $R(H)$ curve was used
which is slightly shifted away from $H=0$ due to the finite coercivity field
resulting in hysteretic $R(H)$ curves. For clarity, in Fig.~\ref{rh} we only
have plotted the data for one sweep direction, i.~e.~the hysteretic behavior
upon field reversal is not shown. Fig.~\ref{rh} shows that the TMR ratio of
the GBJ is drastically reduced with increasing bias voltage. This effect is
clearly found for 4\,K$\le T\le$120\,K, where the GB resistance dominates
and non-ohmic IVCs are observed\cite{comment}.


\vspace*{0cm}
\begin{figure}[h]
\centerline{\mbox{\epsfxsize=0.65\columnwidth
\epsffile{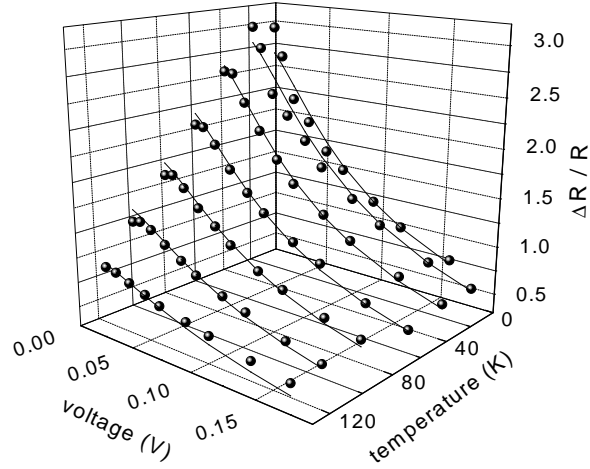}}} \caption[]{\small $\Delta R/R$
(symbols) at $H=1.5$\,T plotted versus bias voltage and
temperature for a 24$^{o}$ [001] tilt LCMO-GBJ. The lines show
$\frac{I^{e}_{ap}}{I^{e}_{ap} + I^{i}}\left( \frac{\Delta R}{R}
\right) _{\rm Julli\grave{e}re}(T)$.  }
 \label{mrv}
\end{figure}


In order to further analyze the data, in Fig.~\ref{mrv} we plotted $\Delta
R/R$ for $\mu_0H=1.5$\,T together with $\frac{I^{e}_{ap}}{I^{e}_{ap} +
I^{i}}\left( \frac{\Delta R}{R} \right) _{\rm Julli\grave{e}re}$  versus the
applied bias voltage. Here, the ratio $\frac{I^{e}_{ap}}{I^{e}_{ap} +
I^{i}}$ has been determined from the measured zero field IVCs using the
GM-model as described above and $\left( \frac{\Delta R}{R} \right) _{\rm
Julli\grave{e}re}(T)$ is chosen to give the optimum fit to the measured
$\Delta R/R$ data. Fig.~\ref{mrv} demonstrates that there is good
coincidence between the two quantities as expected according to the
three-current model. This strongly suggests that the observed decrease of
the TMR ratio of the LCMO-GBJs is due to inelastic tunneling processes
across a GB barrier containing a high density of localized defect
states\cite{Westerburg:x}.

Fig.~\ref{mrv} shows that for $\Delta R/R$ a maximum value of about 300\% is
obtained at low $T$ and $V$. Formally, this corresponds to a spin
polarization of about 80\%.  We note, however, that the actual spin
polarization in LCMO may be slightly larger and the derived value of 80\%
only represents a lower estimate, since we do not know the detailed domain
structure in the LCMO film. Fig.~\ref{mrv} clearly shows that the TMR ratio
strongly decreases with increasing $V$. This behavior is well described
within the three-current model discussed above due to an increase of the
inelastic tunneling current with increasing $V$. The three-current model
provides an intuitive description of the magnetotransport suggesting that
for the manganite GBJs there is a significant inelastic tunneling current
mediated by multi-step tunneling via localized defect states within the GB
barrier. That is, the strong reduction of the GB magnetoresistance with
increasing $T$ and $V$ is caused by an imperfect tunneling barrier
containing a large density of localized defect states. For the LCMO-GBJs
this most likely is related to disorder, strain and oxygen non-stoichiometry
at the GB. Evidently, a considerable improvement of the TMR ratio is
possible by reducing the density of defect states in the tunneling barrier.
To what extent other inelastic processes such as the excitation of surface
magnons play a role cannot be derived from our present data and has to be
further evaluated. We finally note that similar results have been obtained
for $\rm La_{2/3}Sr_{1/3}MnO_3$-GBJs and other misorientation angles showing
that the observed behavior seems to be general for GBJs in the doped
manganites.

In the three-current model the inelastic tunneling channel provides an
additional transport channel that does not depend on the relative
orientation of the magnetization. That is, this inelastic channel acts as a
parallel shunt for the resistance representing the elastic, spin
polarization conserving channel. It is obvious that this results in a
reduction of the TMR ratio. We note that this situation in ferromagnetic
LCMO-GBJs is analogous to that in superconducting cuprate GBJs. For the
latter, phase coherence is lost in the inelastic channels and therefore
Cooper pairs cannot be transferred via these channels in the same way as
spin polarized electrons in magnetic tunnel junctions. For superconducting
tunnel junctions the resistive shunt due to the additional inelastic channel
results in a reduction of the characteristic junction voltage equivalent to
the reduction of the TMR ratio in magnetic junctions. Recently, Gross {\it
et al.}\cite{Gross:97a,Marx:95a} proposed the {\em Intrinsically Shunted
Junction} model to explain the effect of inelastic tunneling on the
properties of GBJs in the cuprate superconductors.

In summary, we have performed a systematic analysis of the $V$ and $T$
dependence of the low field TMR of LCMO-GBJs. We found a strong decrease of
$\Delta R/R$ with increasing $V$ and $T$.  As key result we found that
$\Delta R/R$ is proportional to $I^{e}/(I^{e}+I^{i})$. This gives strong
evidence that the reduction of the TMR ratio is caused by an increase of
$I^{i}$. We have used a three-current model that extends the well known
two-current Julli\`{e}re model by adding an additional inelastic tunneling
contribution. This model naturally explains the observed proportionality and
astonishingly well describes the $V$ dependence of the measured TMR ratio.
Our results suggest that the drastic decrease of the GB-TMR in manganites is
caused by an imperfect tunneling barrier and can be at least partly avoided
by improving the barrier quality. We finally note that the three-current
model should be applicable to other magnetic tunnel junctions or granular
systems where imperfect barriers also may be the origin of a considerable
$V$ and $T$ dependence of the TMR.

The authors thank H. Micklitz for valuable discussions.  This work is
supported by the Deutsche Forschungsgemeinschaft.

\end{document}

%% file: euromacr.tex
\newif\ifboo \boofalse


%% file: Hoefener00EPL.bbl
\begin{thebibliography}{99}
\bibitem{comment}
At $T>120$\,K, the additional series resistance of the film regions adjacent
to the GB is no longer small. Then, the subtraction procedure results in
larger errors and the interpretation of the data is no longer reliable.
\end{thebibliography}
